# Effects of Solar Radio Emissions on Outdoor Propagation Path Loss Models at 60 GHz Bands for Access/Backhaul Links and D2D Communications

Ahmed Iyanda Sulyman, *Senior Member, IEEE*, Hussein Seleem, *Student Member, IEEE*, Abdulmalik Alwarafy, Khaled M. Humadi, and Abdulhameed Alsanie

*Abstract*—This paper presents analytical and empirical data documenting the effects of solar radio emissions on outdoor propagation path loss at 60 GHz bands. Both line-of-sight (LOS) and non-LOS scenarios were considered. The setup used in the empirical studies emulates the future fifth-generation cellular systems for both access and backhaul services, as well as for device-to-device communications. Based on the measurement data collected in sunny weather with intense solar activities, we developed large-scale propagation path loss models at 60 GHz, and observed the effects of solar radio emissions on the path loss data. It is shown that solar radio emission can decrease carrier-to-noise ratio, and that this translates into a corresponding increase in the path loss exponent (PLE) values for the large-scale propagation path loss channel models. Empirical data show that 9.0%–15.6% higher PLE values were observed in hot and sunny weather during the day (41°–42 °C) compared with the counterpart measurements taken at night in cool and clear weather (20°–38 °C). This translates into a corresponding decrease in 60 GHz radio coverage in hot and sunny weather during the day. The empirical data are closely corroborated by analytical estimates presented.

## I. Introduction

MILLIMETER wave (mmWave) spectrum, 30–300 GHz, will play a key role in the fifth-generation (5G) cellular networks, which aim to provide multigigabit per second (Gb/s) data rates over wireless links. Experts have been exploring the prospects of the 28, 38, 60, and 73 GHz mmWave bands for 5G systems. The unlicensed spectrum at 60 GHz offers 10–100 times more spectrum than what is available today for users of 4G cellular systems that operate at frequencies below 6 GHz [1], [2]. The vast amount of unregulated spectrum allocation for 60 GHz devices in many countries (roughly 7 GHz [1]) has attracted the cellular industry to explore the possibility of some harmonizations of this spectrum for use in the 5G mobile networks.

The 60 GHz frequency band has become popular recently through the commercial deployment of consumer devices for wireless local area networks (WLANs) at 60 GHz mmWave band. The three dominant 60 GHz WLAN specifications are Wireless HD, the IEEE 802.11ad (WiGig), and the IEEE 802.15.3c [1]. The economy of scales resulting from the commercial deployments of these schemes has helped reduce the cost of 60 GHz on-chip integrated circuit technologies significantly. This coupled with the increasing demand for high data rate mobile radio services and the huge signal bandwidths available in this band; all have led researchers to consider the use of cellular radio links at the 60 GHz mmWave band. As cellular operators continue to move to smaller cell sizes to exploit spatial reuse, base stations (BSs) will become more densely distributed in urban areas. Therefore, many small cells (microcells and picocells) will be deployed in the 5G cellular networks in urban areas, where low-powered BSs could possibly be mounted on street lamps [3]–[6]. Cost-effective handling of access and backhaul services in such dense network deployments becomes an important consideration. Particularly, the huge number of BSs involved could overstretch the existing backhaul resources. Wireless backhaul in unlicensed mmWave band will thus become an essential part of network design to reduce operating costs. The 60 GHz radio can play key roles in this regard, providing cost-effective short range access and backhaul links in 5G networks that can be harmonized with the operations over licensed bands.

Fig. 1 displays an example of typical 60 GHz wireless link deployment in 5G cellular networks for backhaul service. Use case scenarios for this type of service include data exchanges between microcell BSs and picocell BSs, BSs and relay BSs/repeaters, as well as other multihop data relays services for backhaul applications. The use of 60 GHz wireless link in these cases provides cheap replacements for fiber/copper backhaul, since it can provide multi-Gb/s wireless data in line-of-sight (LOS) communication scenarios using highly directional antennas [1], [7]. Fig. 1 also displays an example of 60 GHz wireless link deployment in 5G cellular networks for access services, where highly directional beams are employed to provide access links to desired nodes while minimizing interference to other nodes in the network.

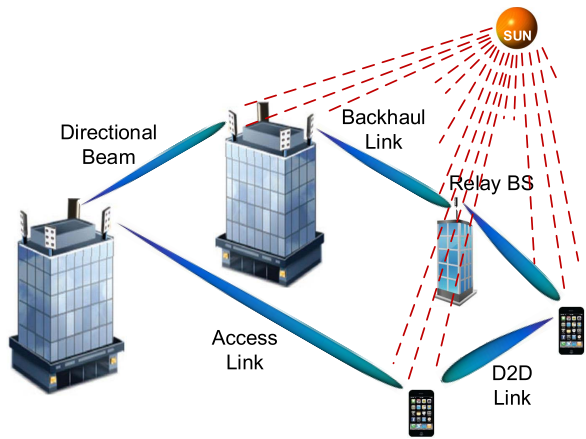

Fig. 1. 60 GHz links deployed in 5G networks for access and backhaul services as well as D2D.

Furthermore, massive device-to-device (D2D) communication links will be deployed in 5G cellular networks. The 3GPP alliance just ratified the narrow-band Internet-of-Things (IoT) standard for cellular networks [8], providing support for low-capacity D2D communications and other low-capacity IoT applications. Support for bandwidth-intensive IoT systems, such as those being developed under the IoT-A reference model [9], is expected later in the context of the 5G cellular standard. All these efforts are expected to enable massive D2D communications in the cellular networks, and 60 GHz radio can provide cost-effective D2D links over the unlicensed bands, as shown in Fig. 1.

Successful deployments of 60 GHz links in cellular networks worldwide, however, require prior real-field measurement campaigns in different regions. This will allow cellular system developers to explore different channel characteristics at this band in different parts of the world. To date, there are significant measurement works at 60 GHz band characterizing the effects of oxygen and water vapor absorptions as well as other terrestrial channel impairments. For example, atmospheric absorption at 60 GHz has been documented in the literature to be about 20 dB/km (or 1.4 dB at 70 m), while attenuation due to heavy rain is about 10 dB/km (or 0.7 dB at 70 m) [1]. Previous works on 60 GHz wireless links in [10] and [11] were focused on indoor propagations for WLAN applications, where the statistical properties of the LOS propagation channels at 60 GHz were studied for various indoor corridor environments. Several other works have also explored other characteristics of the 60 GHz mmWave propagation channel. In [12], it was shown analytically that outdoor mesh networks based on short-range 60 GHz links can be used to provide a multigigabit wireless backhaul for picocells, or for last-hop links in a neighborhood. The work in [3] explored the multipath characteristics of mmWave channel at 60.4 GHz, based on LOS propagation measurements conducted in a corridor and square empty room indoors using omnidirectional antennas. Early works in [13] found that it is possible to establish 60 GHz cellular radio links in short-range point-to-point scenarios, based on LOS mmWave propagation measurements conducted outdoors using electronic beam steering techniques. Recent works in [4] and [5] developed large-scale and small-scale channel parameters for cellular systems at 60 GHz, based on LOS measurement campaigns conducted in a busy urban outdoor environment using omnidirectional antennas. The measurement environments in those works emulate typical small cell deployment scenarios in heterogeneous networks. The work in [14] considered military applications in a forest environment, and characterized the severity of multipath in an LOS mobile radio microcellular wireless channel in the presence of foliage. They used omnidirectional antennas at 60 GHz. Ben-Dor et al. [15] conducted wideband propagation measurements at 60 GHz in an outdoor campus environment for both LOS and non-LOS (NLOS) propagation scenarios. They obtained the corresponding large-scale path loss models for cellular peer-to-peer (or D2D) communications. The work in [16] considered both narrow-band and wideband measurements. They obtained empirical data for various outdoor environments (such as airport field, urban street, and city tunnel), which are useful for the development of prediction models at 60 GHz in different outdoor propagation environments. Violette et al. [17] studied the characteristics of mmWave signals when propagating through building materials and other structures commonly found in a city, based on measurement works conducted in an urban environment considering both LOS and NLOS propagation scenarios. Despite all these works, however, there are only few measurement works currently available on the effects of solar radio emissions at 60 GHz band.

Recently, Sulyman et al. [2] presented the initial results documenting the effects of solar radio emissions on path loss exponent (PLE) values at 60 GHz band based on outdoor measurements conducted in an urban environment. Solar radio noise is a phenomenon well known in the satellite community [18], [19], [26], [30], [32], [33]. In the context of terrestrial wireless communications, however, it was examined only recently [20], [27], [28], [31]. To the best of our knowledge, this paper presents the first comprehensive analysis of the effects of solar radio emissions on 60 GHz mmWave path loss, using measurement data collected in the gulf region where extreme temperatures and intense solar radiation are encountered. The effects of hot and sunny weather conditions on the PLE values at 60 GHz bands are then carefully estimated. Our empirical data indicate significant increase in path loss in hot and sunny weather in Riyadh City during the day, compared with a cool and clear weather at night. Analytical expressions are also provided for evaluating the decrease in carrier-to-noise ratio (CNR) due to solar radio emissions, and the analytical results corroborate the empirical data very closely. Even though the empirical results in this paper were collected in the gulf region, the observed effects of solar radio emissions on the path loss are applicable in any region of the globe that experiences intense solar activities.

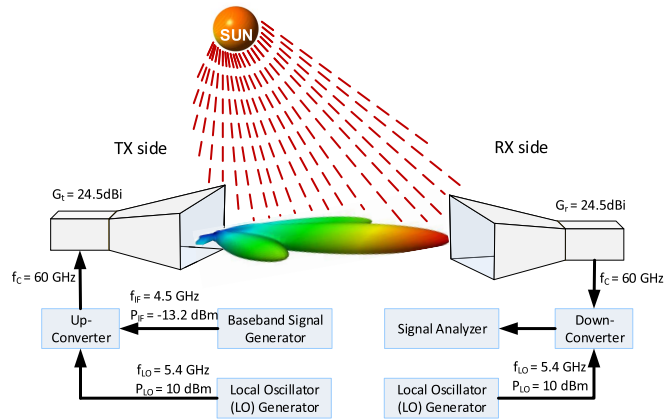

Fig. 2. Block diagram of the hardware used for the 60 GHz mmWave propagation channel measurements conducted in Riyadh City.

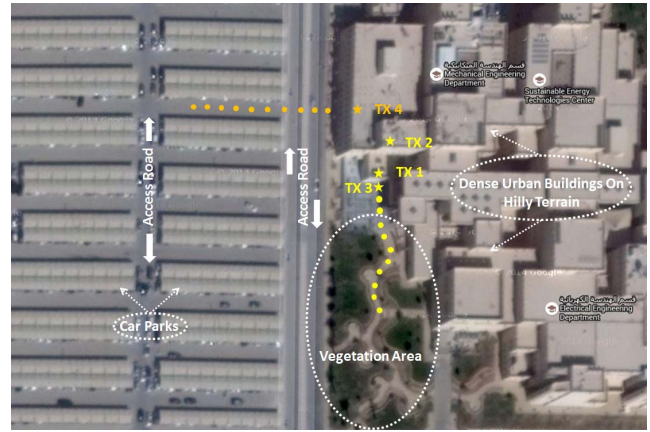

Fig. 3. Outdoor environments at the college of Engineering building, KSU campus in Riyadh City, where we considered four measurement scenarios with TX heights TX 1, TX 2, TX 3, and TX 4. The big stars represent TX locations while the circles represent RX locations.

The rest of this paper is organized as follows. Section II describes the measurements setup and environments used in the propagation measurements. Section III presents the analysis of the effects of solar radio emissions on path loss channel models at 60 GHz. Section IV presents the results, while Section V summarizes the conclusions of this paper.

## II. EXPERIMENTAL PROCEDURE

### A. Measurement Setup

In our measurement setup, we model 60 GHz wireless links in the future 5G cellular network by transmitting a 60 GHz carrier signal and measuring the received power at various transmitter–receiver (TX–RX) separation distances. Then, we obtained the large-scale propagation path loss channel parameters from the empirical data collected. The measurements were conducted using a continuous wave (CW) signal transmitted from a fixed TX to a mobile RX, and the received signal power was recorded as a function of mobile positions. The 60 GHz mmWave CW signal is generated using the hardware shown in Fig. 2. At the TX side, a vector signal generator was used to internally generate a sinusoidal signal at an intermediate frequency of 4.5 GHz (i.e., $f_{\text{IF}} = 4.5$ GHz). This signal was mixed up inside the upconverter with a local oscillator (LO) signal that was generated from an analog signal generator at a frequency of 5.4 GHz (i.e., $f_{\text{LO}} = 5.4$ GHz). The output of the mixer is a radio frequency (RF) signal at 60 GHz mmWave band (i.e., $f_c = 60$ GHz) [2]. The RF signal at the output of the mixer was then fed onto a highly directional horn antenna [with 7.3° half-power beamwidth (HPBW)], and finally transmitted over the wireless channel. At the RX side, an identical horn antenna was used to capture the 60 GHz RF signal. The received 60 GHz signal was downconverted using an identical LO signal at 5.4 GHz, in order to retrieve the original message at 4.5 GHz. This signal was finally fed onto a spectrum analyzer from where the received power levels were measured. The rest of the measurement parameters that were used in our measurement campaigns are listed in Table II.

### B. Measurement Environments

The mmWave measurement campaign presented here was conducted in three distinct propagation environments. One of these is an indoor environment, while the other two are outdoor environments. The indoor measurement was used only as a baseline assessment, to double-check the accuracy of the data collected outdoors later on in the campaign using the same equipment. The indoor environment was chosen inside the college of Engineering building at King Saud University (KSU), Riyadh, Saudi Arabia, while the outdoor environments were chosen in different areas around KSU campus.

The indoor power measurement data were collected in a long hallway, dimensions 160 m × 7.15 m × 12 m, with a number of doorways leading outside the building. The walls in this environment were made of tinted glass and concrete materials. The ceilings were concrete, and the ground was covered with marble tiles. Along the whole length of the hallway, there were a number of concrete pillars spaced uniformly [2].

The outdoor measurement campaigns were conducted in two distinct urban outdoor environments around KSU campus in Riyadh City, in the periods between the summer of 2015 and fall of 2016. Both LOS and NLOS propagation scenarios were considered, and the measurement setups were chosen carefully to emulate access links between microcells/picocells BSs and mobile devices, backhaul links for micro/picocells BSs, as well as D2D communication links between mobile devices. The first outdoor urban cellular environment was a hilly terrain type with high-rise buildings and some vegetation as shown in Fig. 3, while the second environment represents a hilly terrain type with high-rise buildings and rare vegetation [2]. In the first environment, we considered two 60 GHz deployment scenarios, access and backhaul links, while in the second environment, we considered two other scenarios, access and D2D links. Each of these scenarios was modeled using appropriate combinations of TX–RX antenna heights. In the first outdoor environment with some vegetation area, we first considered a deployment scenario, which models cellular backhaul links. The TX height used was 8.5 m (i.e., $h_{\text{tx}} = 8.5$ m) and depicted as TX 1 in Fig. 3, while the RX height was 3.5 m (i.e., $h_{\text{rx}} = 3.5$ m). Mobile RX with heights beyond 3.5 m could not be used due to practical constraints, but this setup still reasonably models some backhaul links for mirco/picocells. In this scenario, up to 15 locations were selected for the power measurements, with TX–RX separation distances ranging from 1 up to 90 m. In the second outdoor deployment scenario in this environment, the TX height used was 18 m (i.e., $h_{\text{tx}} = 18$ m) and depicted as TX 2 in Fig. 3 while the RX height was 3.5 m (i.e., $h_{\text{rx}} = 3.5$ m). This setup models an access link scenario. In this scenario, up to 11 locations were selected with TX–RX separation distances ranging from 1 up to 80 m. In the second outdoor environment representing a rare vegetation area, we first considered an access link deployment scenario where the TX height was 14 m (i.e., $h_{\text{tx}} = 14$ m) and depicted as TX 3 in Fig. 3 while the RX height was 2 m (i.e., $h_{\text{rx}} = 2$ m). In this scenario, ten measurement locations were selected with TX–RX separation distances ranging from 1 up to 110 m. In the second outdoor deployment scenario in this environment, we modeled D2D communication links. D2D communication links were established between the TX and the RX using TX and RX antenna heights of 1.9 m. This case models a scenario in the future 5G cellular network whereby mobile devices interact with each other over the 60 GHz radio, and the aggregated data possibly eventually transmitted over the cellular network via network access provided to any of the interacting devices [29]. At each and every outdoor measurement location, the TX antenna was manually downtilted toward the RX and then fixed in a location. Then, the RX was moved from one location to another. The RX antenna was manually adjusted in such a way to get the best wireless communication link, as observed from the readings on the spectrum analyzer, at each and every location. Also, the spectrum analyzer and the signal generators were all kept under the shade provided by the mobile trolley used in the measurements in order to minimize the effects of temperature drifts.

## III. Effects of Solar Radio Emissions on mmWave Path Loss Models at 60 GHz

The directional propagation path loss channel models developed from the measurement campaigns described in Section II are presented in this section, using the log-distance and the floating intercept modeling approaches. The path loss is expressed in dB as a function of TX–RX separation distance on a logarithmic scale.

### A. Log-Distance Path Loss Models

The log-distance path loss model (LDM) is given by [1], [23]

$$\text{PL}(d)[\text{dB}] = \text{PL}(d_0)[\text{dB}] + 10n \log_{10}\left(\frac{d}{d_0}\right) + X_\sigma \quad d \geq d_0 \quad (1)$$

where $\text{PL}(d_0)[\text{dB}]$ is the path loss in dB at a reference distance $d_0$, $n$ is the PLE, and $X_\sigma$ is a zero-mean Gaussian variable in dB (log-normal) with standard deviation $\sigma$ in dB (also called shadow factor). Throughout this paper, we used $d_0 = 1$ m, and $\text{PL}(d_0)[\text{dB}]$ can be estimated using Frii's free space path loss equation

$$\text{PL}(d_0)[\text{dB}] = 20 \log_{10}\left(\frac{4\pi d_0}{\lambda}\right) \quad (2)$$

where $\lambda$ is the carrier signal wavelength in meters. The parameters of this model are then obtained by finding the best minimum mean square error line fit to the empirical data collected during measurement campaigns.

### B. Floating Intercept Path Loss Models

The floating intercept path loss model (FIM) is typically used in the 3GPP sets of models, and it is given by [24]

$$\text{PL}(d)[\text{dB}] = \alpha + 10\beta \log_{10}(d) + X_\sigma \quad (3)$$

where $\alpha$ is the floating intercept in dB and $\beta$ is the linear slope. The parameters of this model are found by applying the least-square linear regression fit to the empirical data in order to find the best-fit values for $\alpha$ and $\beta$ with minimum standard deviation [24]. Note that with large enough measurement data taken, the path loss curves from the two models in (1) and (3) will converge, providing some means of assessing the degree of accuracy of the measurement data collected.

### C. Effects of Solar Radio Emissions on 60 GHz Radio Links

The sun is a wideband radio emitter whose radio emissions manifest during radio wave detections in a similar way as the Gaussian noise. However, interference due to solar radio emissions is quantifiable, and they are frequency-dependent, unlike the Gaussian noise, which is entirely random and frequency-independent. Solar radio emission is more noticeable in the detection of weak signals [18], [21], such as the satellites and the 60 GHz radio signals. The higher the intensity of solar activities in an area, the more noticeable the associated radio interference due to solar emissions in an outdoor radio signal detection in the area. Solar radio emissions consist of three components: the background or quiet component, the slowly varying component, and the burst component. The quiet component comes mostly from the featureless areas of the solar disc, the slowly varying component comes from the light white area of the sun (known as the Plage), while the burst component comes from solar flares in the intensely bright area of the sun. Solar radio emissions are typically expressed in solar flux unit (SFU), with one SFU equaling $10^{-22}$ W m$^{-2}$ Hz$^{-1}$.

The integrated flux density, $S$, due to all solar radio emissions at 60 GHz can be estimated using the expression

$$S = S_q + S_v + S_b \quad (4)$$

where $S_q$, $S_v$, and $S_b$ denote the quiet, the slowly varying, and the burst components of the solar radio emissions, respectively. $S_q$ is a fixed component at a particular frequency, $S_v$ varies over time for any given frequency, while $S_b$ is a fixed number (0 or $B$), where $B$ is the burst value whenever solar flare exists during signal detections.

An approximate analytical estimate for $S_q$ at 60 GHz can be obtained using the expression [18]

$$S_q(f_{\text{GHz}}) \approx 26.4 + 12.4 f_{\text{GHz}} + 1.11 f_{\text{GHz}}^2 \text{ SFU} \quad (5)$$

where $f_{\text{GHz}}$ is the carrier frequency expressed in GHz. This expression is accurate within 10% in the 1–20 GHz frequency range, but it becomes more exaggerating at higher frequencies [18]. For this reason, we have used it here only as an approximate analytical estimate for $S_q$ at 60 GHz. The slowly varying component $S_v$ can be estimated accurately, within 10% accuracy, at any frequency using the expression [18]

$$S_v(f_{\text{GHz}}) \approx \frac{0.64(F_{10} - 70) f_{\text{GHz}}^{0.4}}{\left(1 + 1.56\left(\ln\left(\frac{f_{\text{GHz}}}{2.9}\right)\right)^2\right)} \text{ SFU} \quad (6)$$

where $F_{10}$ is the current value of the solar radio emission measured at a wavelength of 10.7 cm or 2.8 GHz frequency. A large value of $F_{10}$ indicates an overall intense solar activities at that time. Observations at the Algonquin Radio Observatory, ON, Canada, shows that $S_v(2.8)$ varies slowly between 0 and over 250 SFU for some 11 years period as overall solar activities waxes and wanes [18]. Solar activity observatory systems, such as the setup in [22], capture all stable solar radio emissions at the wavelength of the measurement. Therefore, an observed $F_{10}$ represents the overall stable solar activity at 2.8 GHz. This means that by definition, $F_{10} = S_q(2.8) + S_v(2.8)$. Notice from (6) that when $F_{10} = 70$ SFU, we obtain $S_v = 0$ SFU, and the value of $S_q$ at that moment is accurately estimated as $S_q(2.8) = F_{10}$. At 2.8 GHz, the value of $S_q$ given by (5) is $26.4 + 12.4(2.8) + 1.11(2.8^2) = 69.8$ SFU, confirming the accuracy of (5) at lower frequencies. Analytical expressions are not readily obtainable for the burst component, $S_b$, but at 20 GHz, the average maximum flux density observed due to the largest burst is 10 000 SFU [18]. The average maximum flux density due to burst is even lower than this value at 60 GHz, but for an upper-bound estimate, it suffices to use 10 000 SFU for this component whenever solar flare is present during 60 GHz radio signal detection, or use 0 SFU otherwise. In most practical applications, $S_b = 0$ is used [18]. For this reason, we have used $S_b = 0$ throughout this paper. Fig. 4 shows the components as well as the integrated flux density from solar radio emissions at different frequencies. It is clear from Fig. 4 that the integrated flux density increases with frequency despite the fact that the varying component has a peak value around 3 GHz and falls off at higher frequencies.

In general, the decrease in CNR ($\Delta$CNR) due to solar radio emission can be estimated using the equality

$$\Delta\text{CNR[dB]} = 10 \log_{10}\left(\frac{C}{N_{kTB}}\right) - 10 \log_{10}\left(\frac{C}{(N_{kTB} + N_{\text{sun}})}\right)$$
$$= 10 \log_{10}\left(\frac{N_{kTB} + N_{\text{sun}}}{N_{kTB}}\right) \quad (7)$$

where $C$ denotes the carrier signal level, while $N_{kTB}$ and $N_{\text{sun}}$ denote, respectively, the thermal noise and solar radio emissions encountered during signal propagations and detections. For single-tone measurements with signal bandwidth $B = 1$, and for narrow beamwidth antennas where the HPBW is less than the visible solar disc diameter, the decrease in CNR ($\Delta$CNR) due to solar radio emissions can be estimated as [18]

$$\Delta\text{CNR[dB]} = 10 \log_{10}\left(1 + \frac{2\pi S A_e}{GkT\Omega_s}\right) \quad (8)$$

where $A_e = (\lambda^2/4\pi)G$ is the antenna aperture in meter square, $G$ denotes the antenna gain, $T$ is the system temperature in degree kelvin (K), $k = 1.38 \times 10^{-23}$ J/K is Boltzmann's constant, and $\Omega_s = 6.8 \times 10^{-5}$ steradian denotes the mean solid angle subtended by the sun.

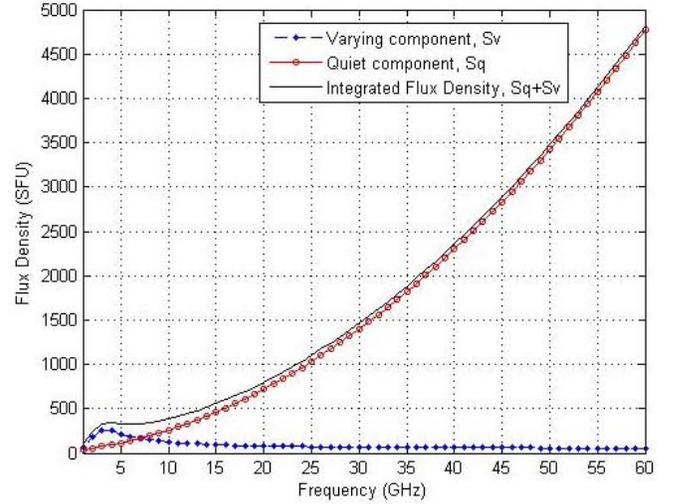

Fig. 4. Flux density from solar emissions as a function of frequency for $F_{10} = 329$ SFU, which corresponds to the peak value of $S_v(2.8) = 250$ SFU recorded at 2.8 GHz in [18].

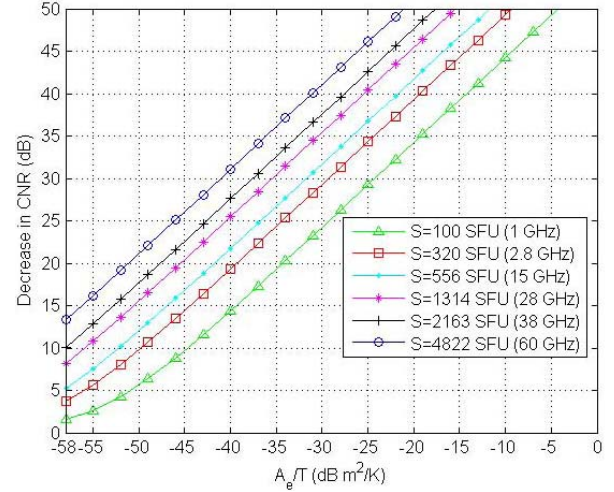

Fig. 5. Analytical estimate for decrease in outdoor CNR ($\Delta$CNR) due to solar radio emissions as a function of antenna aperture-to-temperature ratio ($A_e/T$). Notice that $\Delta$CNR $= 13.3$ dB at 60 GHz when $A_e/T = -58$ dB m$^2$/K.

Fig. 5 shows the $\Delta$CNR curves obtained from (4)–(8) as a function of antenna aperture-to-temperature ratio ($A_e/T$), for various carrier frequencies. It is easily observed from Fig. 5 that for the 60 GHz links [$S = 4822$ SFU from (4)–(6)], $\Delta$CNR $= 13.3$ dB at $A_e/T = -58$ dB m$^2$/K, which corresponds to 24 dBi antenna used in our outdoor measurements at 42 °C. Analytical estimates for the received signal power $P_{rec}(d)$[dB] at different TX–RX separation distances $d$ can be obtained using

$$P_{rec}(d)[\text{dB}] = P_t[\text{dB}] - PL(d)[\text{dB}] + G_t[\text{dB}] + G_r[\text{dB}] \quad (9)$$

where $P_t$[dB] is the transmitted power in dB, PL($d$)[dB] is given by (1), while $G_t$[dB] and $G_r$[dB] are, respectively, the TX and RX antenna gains in dB. Table I documents the analytical CNR given by: CNR = $P_{rec}(d)/kTB$ for various TX–RX separation distances $d$, and the percentage degradation represented by $\Delta$CNR = 13.3 dB. It is observed from Table I that on average, 9.6% degradation in CNR is expected over 100 m TX–RX separation distance when using 60 GHz single-tone channel sounder in a 42 °C sunny weather, with $P_t$[dB] = 10 dBm and 24 dBi antennas (corresponding to $A_e/T = -58$ dB m$^2$/K). This will result in a corresponding degradation in the PLE values, since PLE by definition represents the average carrier signal attenuation (in dB), per decade of log TX–RX separation distance. Our measurement data presented later agree closely with this theoretical estimate.

TABLE I
ANALYTICAL % CNR DEGRADATIONS DUE TO SOLAR RADIO EMISSIONS FOR 60 GHz LINKS USING 10 dBm TX POWER AND 24 dBi ANTENNAS IN 42 °C SUNNY WEATHER, OR $A_e/T = -58$ dB m$^2$/K, CORRESPONDING TO $\Delta$ CNR = 13.3 dB USING (4)–(8)

| TX-RX separation distance $d$ [m] | 1 | 10 | 20 | 50 | 100 |
|---|---|---|---|---|---|
| $P_{rec}(d)$ [dB] (for LOS, PLE=2, $P_t$=10 dBm) | −40.0 | −60.0 | −66.0 | −74.0 | −80.0 |
| CNR [dB] =$10\log_{10}\left(\frac{P_{rec}(d)}{kTB}\right)$ | 163.6 | 143.6 | 137.6 | 129.6 | 123.6 |
| % CNR degradations due to solar radio emissions for $\Delta$ CNR=13.3 dB | 8.13% | 9.26 % | 9.67 % | 10.26 % | 10.76 % |
| Cumulative average over Distance $d$ | 8.1% | 8.7 % | 9.0 % | 9.3 % | 9.6 % |

## IV. MEASUREMENT RESULTS AND ANALYSIS

The path loss models for all the indoor and outdoor environments are presented here using the LDM expression in (1) and the FIM expression in (3). The parameters of the LDM were obtained, with respect to 1 m reference distance, for LOS and NLOS propagation scenarios, and the results are summarized in Table II, where the empirical percentage PLE degradation due to solar radio emissions is presented for outdoor environments, both LOS and NLOS use case scenarios.

### A. Indoor Measurements

The results for indoor measurements are shown in Fig. 6. These results are included here only as a reference, since the indoor measurement data do not contain the effects of solar radio emission under investigation in this paper. However, the fact that these data agree with established concepts is a confirmation of the accuracy of the empirical data collected outdoors later using the same equipment. It is evident from Fig. 6 that the PLE values for LOS scenarios in the indoor measurements are approximately identical to that of free space (i.e., $n \approx 2$), agreeing with Frii's free space equation as expected. It is also noticed from these results that the path loss curves for the LDM and FIM converge fairly well for both the LOS and NLOS scenarios, again confirming the accuracy of the empirical data. In the rest of this paper, we focus, henceforth, on the LDM, because it has a well-established theoretical basis in the literature [1].

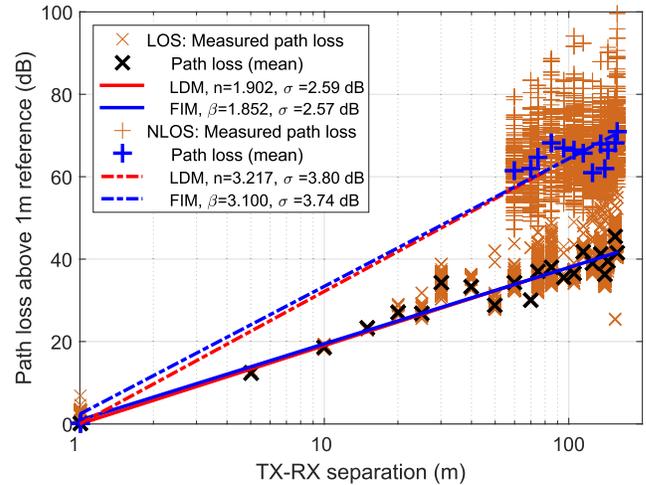

Fig. 6. LDM and FIM along with the measurement data collected at 60 GHz in Riyadh City for an indoor long hallway ($h_{tx} = 1.5$ m and $h_{rx} = 1.5$ m).

### B. Outdoor Measurements

The results for outdoor measurements are summarized in Figs. 7–15, where we have compared outdoor night-time measurements (20°–38 °C) with day-time measurements in sunny and hot weather (41°–42 °C).

*1) Effects of Temperature Drifts:* Temperature drift causes slight increase in noise floor when equipment becomes hot, which can reduce CNR. At room temperature (25 °C), thermal noise power over $B$ bandwidth can be calculated as

$$N_{25°} = 10\log_{10}(kTB) = -173.9 \text{ dBm} + 10\log_{10}(B) \quad (10)$$

whereas at 42 °C the thermal noise power will drift to

$$N_{42°} = 10\log_{10}(kTB) = -173.6 \text{ dBm} + 10\log_{10}(B) \quad (11)$$

which results only in a small change in CNR. Notice that equipment noise figure was omitted in these calculations, because it has the same value in day-time and night-time operations. Also, note that temperature drifts in the baseband and RF components are all estimated using the equations mentioned earlier, and the cumulative effect of these drifts on the system is obtained as the numerical summation of the drifts in the individual units in the communication chain. However, if the same drifts existed in the 1 m reference data collected, then referencing the path loss at different locations to the 1 m data removes the effects of these drifts from the reported path loss above 1 m reference. This is one of the many benefits of presenting path loss above the close-in reference path loss, as we have done in this paper. The remnant effect of temperature drifts that may be left in such cases will be when the equipment becomes slightly hotter at other locations than they were at the 1$m$ reference, which is very minimal as shown in (10) and (11). Thus, temperature drifts cannot be ruled out, but the effect of solar radio emissions appears to dominate, as illustrated in the following.

TABLE II
PERCENT INCREASE IN PLE VALUES IN EXTREME SUNNY WEATHER, WITH RESPECT TO A 1 m REFERENCE DISTANCE, FOR THE LOS AND NLOS OUTDOOR MEASUREMENTS AT 60 GHz IN RIYADH CITY

| Frequency (GHz) | 60 | |
|---|---|---|
| TX gain (dBi) | 24 | |
| RX gain (dBi) | 24 | |
| TX HPBW (°) | 7.3 | |
| RX HPBW (°) | 7.3 | |
| $d_0$ (m) | 1 | |
| TX Power (dBm) | 10 | |

| Scenario: (Outdoor D2D Links) | LOS | NLOS |
|---|---|---|
| Environment | rare vegetations | |
| TX height (m) | 1.9 | |
| RX height (m) | 1.9 | |
| PLE, Sunny Sky, 42°C, day time | 2.559 | 4.219 |
| $\sigma$ (dB) | 1.02 | 3.03 |
| PLE, Night time, 38°C | 2.239 | 4.141 |
| % PLE increase in Sunny Sky | 14.3 % | 1.9 % |
| $\sigma$ (dB) | 1.97 | 2.87 |

| Scenario: (Outdoor Back-haul Links) | LOS | NLOS |
|---|---|---|
| Environment | Hilly with some veget. | |
| TX height (m) | 8.5 | |
| RX height (m) | 3.5 | |
| PLE, Sunny Sky, 41°C | 2.227 | 3.656 |
| $\sigma$ (dB) | 3.89 | 3.64 |
| PLE, Clear night Sky, 20°C | 2.018 | 3.443 |
| $\sigma$ (dB) | 2.42 | 1.75 |
| % PLE increase in Sunny Sky | 15.6 % | 1.5 % |
| PLE, Day-time, 25°C | 2.103 | 3.681 |
| $\sigma$ (dB) | 3.39 | 4.81 |
| PLE, Night-time, 20°C | 1.927 | 3.601 |
| % PLE increase in Sunny Sky | 9.1 % | 2.2 % |
| $\sigma$ (dB) | 3.43 | 4.13 |
| PLE, Dusty Sky, visibility = 3km, 32°C | 2.086 | 3.785 |
| $\sigma$ (dB) | 2.51 | 4.44 |

| Scenario: (Outdoor Access Links) | LOS | NLOS |
|---|---|---|
| Environment | rare veget. | |
| TX height (m) | 14 | |
| RX height (m) | 2 | |
| PLE, Sunny Sky, 41°C | 2.107 | 3.638 |
| $\sigma$ (dB) | 2.94 | 2.77 |
| PLE, Clear night Sky, 20°C | 1.854 | 3.263 |
| % PLE increase in Sunny Sky | 13.7 % | 11.5 % |
| $\sigma$ (dB) | 1.42 | 3.04 |

| Scenario: (Outdoor Access Links) | LOS | NLOS |
|---|---|---|
| Environment | Hilly with some veget. | |
| TX height (m) | 18 | |
| RX height (m) | 3.5 | |
| PLE, Sunny Sky, 41°C | 2.199 | 3.537 |
| $\sigma$ (dB) | 3.71 | 2.18 |
| PLE, Clear night Sky, 30°C | 2.017 | 3.435 |
| % PLE increase in Sunny Sky | 9.0 % | 3.0 % |
| $\sigma$ (dB) | 1.15 | 2.91 |

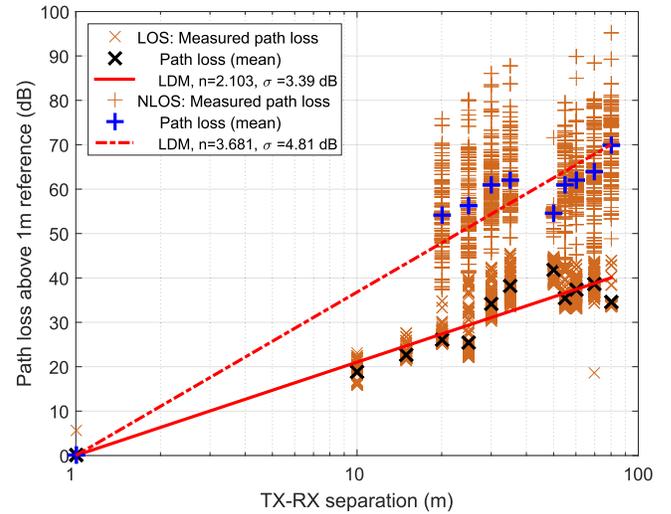

Fig. 7. LDM along with the measurement data collected at 60 GHz in Riyadh City for outdoor backhaul links in urban environment with hilly terrain type and some vegetation in 25 °C day-time weather ($h_{tx}$ = 8.5 m and $h_{rx}$ = 3.5 m).

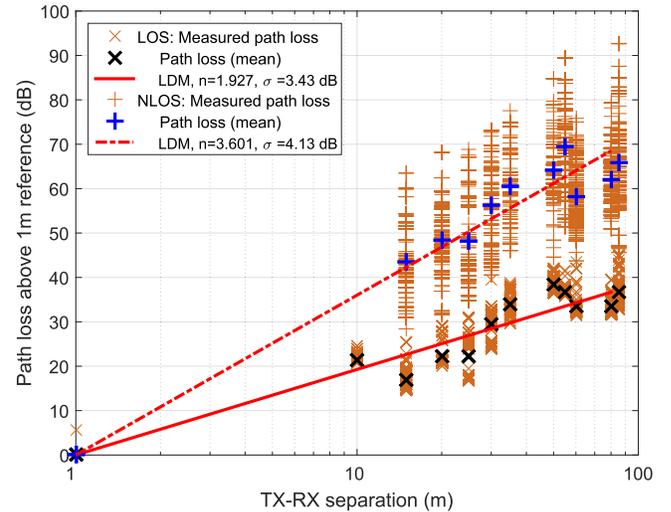

Fig. 8. LDM along with the measurement data collected at 60 GHz in Riyadh City for outdoor backhaul links in urban environment with hilly terrain type and some vegetation in 20 °C night-time weather ($h_{tx}$ = 8.5 m and $h_{rx}$ = 3.5 m).

*2) Effects of Solar Radio Emissions:* The assertion that solar radio emissions result in lower CNR during the day compared with night time was examined further in our measurements. For the outdoor environments, we first collected measurement data for day-time and night-time deployments on a day when system temperatures were very similar during the day (25 °C) and at night (20 °C). These results are summarized in Figs. 7 and 8. In this case, the PLE value in LOS scenario is approximately 9.1% higher during the day compared with the night, and this 9.1% PLE degradation is entirely attributable to the solar radio emission effects.

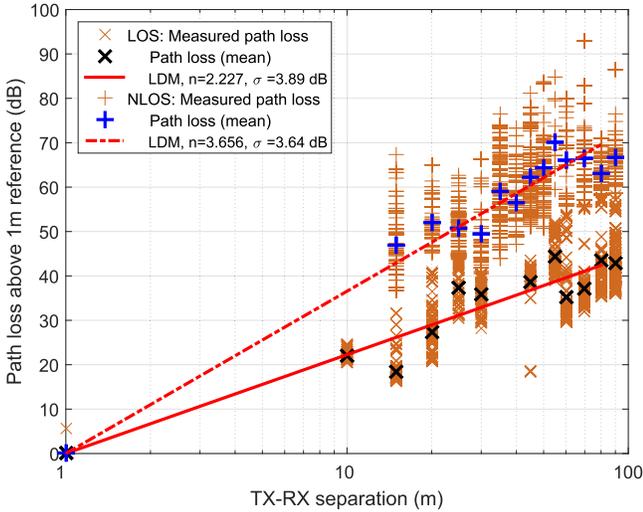

Fig. 9. LDM along with the measurement data collected at 60 GHz in Riyadh City for outdoor backhaul links in urban environment with hilly terrain type and rare vegetation in 41 °C sunny sky ($h_{tx}$ = 8.5 m and $h_{rx}$ = 3.5 m).

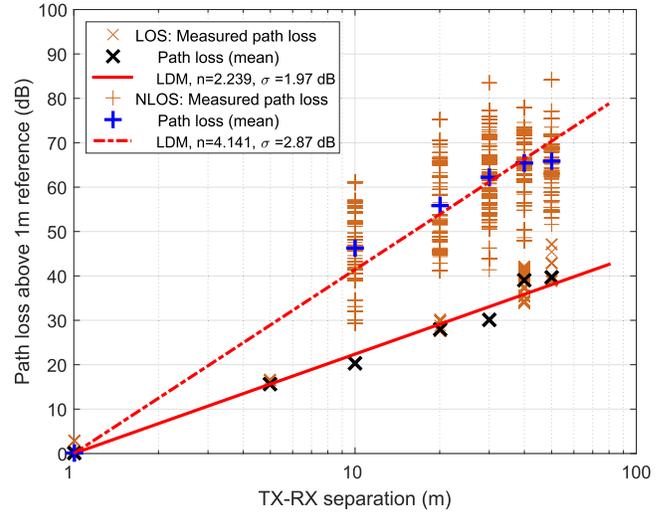

Fig. 11. LDM along with the measurement data collected at 60 GHz in Riyadh City. The measurements were conducted to model D2D communication links in urban environment with rare vegetation in a 38 °C clear night sky ($h_{tx}$ = 1.9 m and $h_{rx}$ = 1.9 m).

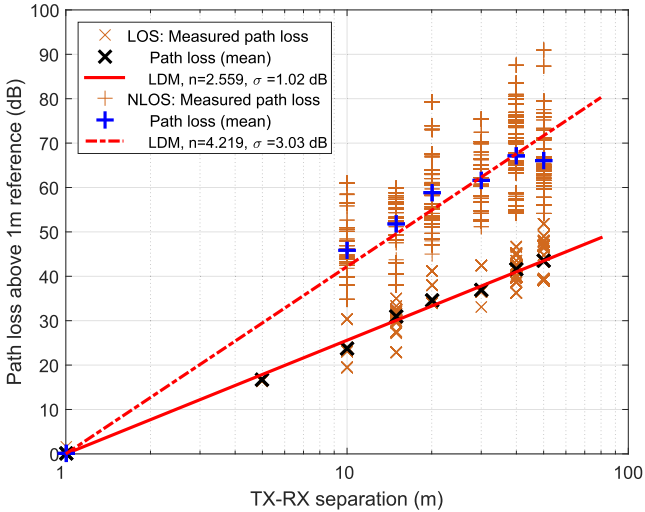

Fig. 10. LDM along with the measurement data collected at 60 GHz in Riyadh City. The measurements were conducted to model D2D communication links in urban environment with rare vegetation in a 42 °C clear sunny sky ($h_{tx}$ = 1.9 m and $h_{rx}$ = 1.9 m).

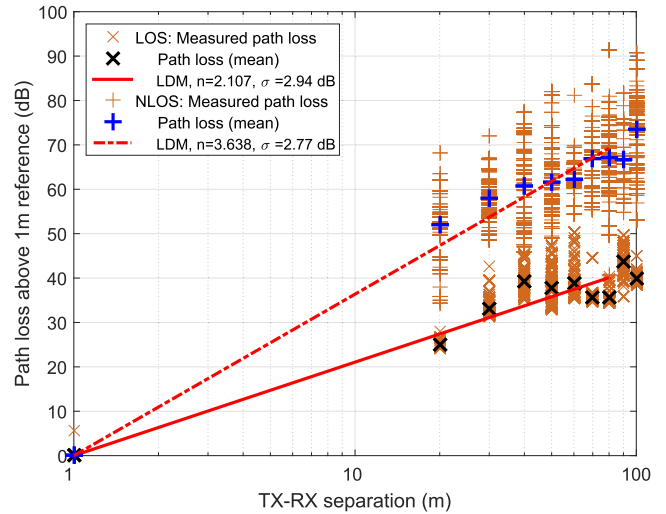

Fig. 12. LDM along with the measurement data collected at 60 GHz in Riyadh City for outdoor Access links in urban environment with hilly terrain type and rare vegetation in 41 °C clear sunny sky ($h_{tx}$ = 14 m and $h_{rx}$ = 2 m).

Next, measurement data were collected during the day in sunny summer afternoons in Riyadh City in 2015 and 2016, with temperature peaking at 41°–42 °C. These data were analyzed and compared with another sets of data collected in clear sky at nights in the fall periods in 2015 and 2016, with temperatures in the range 20°–38 °C. The path loss data obtained are shown in Figs. 9–16. Table II presents a summary of the PLE values obtained from these measurements for cases modeling access, backhaul, and D2D communications. The PLE values are higher in all cases during the day compared with the night time. For the LOS scenarios, roughly 9.0%–15.6% higher PLE values were obtained in sunny summer afternoons compared with their night-time counterparts for measurements modeling access links ($h_{tx}$ = 18 m and $h_{rx}$ = 3.5 m, $h_{tx}$ = 14 m, and $h_{rx}$ = 3.5 m), backhaul links ($h_{tx}$ = 8.5 m and $h_{rx}$ = 3.5 m), and D2D communications ($h_{tx}$ = 1.9 m and $h_{rx}$ = 1.9 m). These results agree closely with the 9.6% theoretical estimate presented in Table I for the cumulative average CNR degradation obtained over similar TX–RX separation distance used in these measurements. For the NLOS case, the PLE values during the day were roughly 11.5% higher than night-time counterparts for access links ($h_{tx}$ = 14 m and $h_{rx}$ = 3.5 m), 1.5% higher for backhaul links ($h_{tx}$ = 8.5 m and $h_{rx}$ = 3.5 m), and 1.9% higher for

6632　　IEEE TRANSACTIONS ON ANTENNAS AND PROPAGATION, VOL. 65, NO. 12, DECEMBER 2017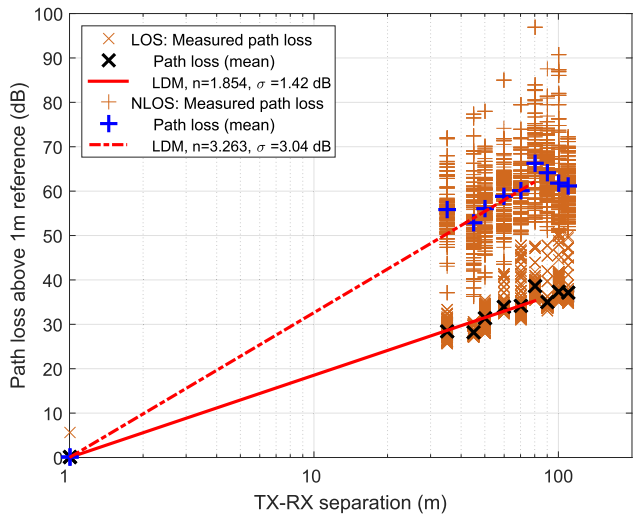

Fig. 13. LDM along with the measurement data collected at 60 GHz in Riyadh City for outdoor Access links in urban environment with hilly terrain type and rare vegetation in 20 °C clear night sky ($h_{tx} = 14$ m and $h_{rx} = 2$ m).

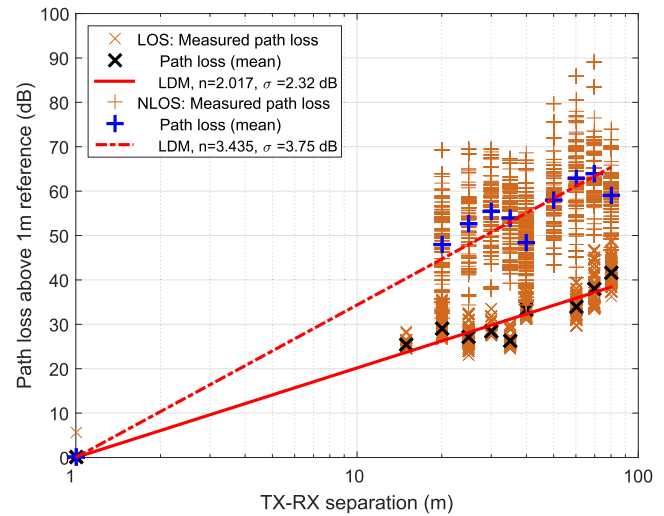

Fig. 15. LDM along with the measurement data collected at 60 GHz in Riyadh City for outdoor access link scenario in 30 °C clear night sky ($h_{tx} = 18$ m and $h_{rx} = 3.5$ m).

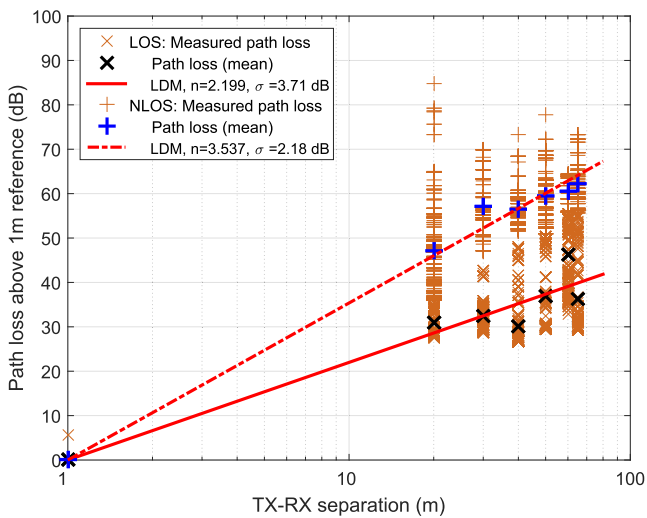

Fig. 14. LDM along with the measurement data collected at 60 GHz in Riyadh City for outdoor access link scenario in 41 °C sunny sky ($h_{tx} = 18$ m and $h_{rx} = 3.5$ m).

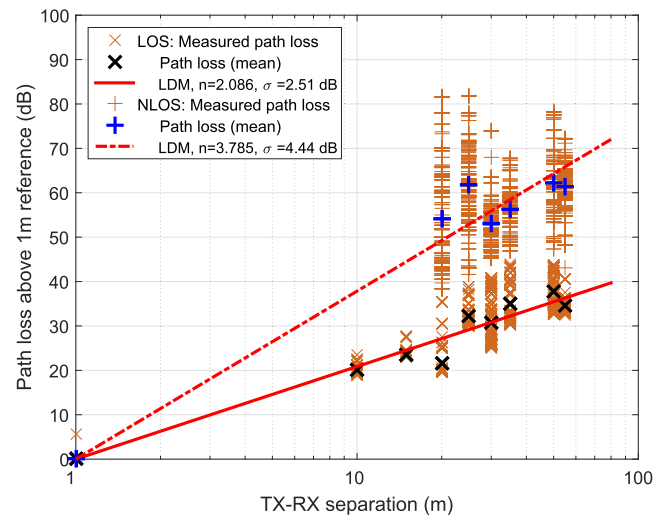

Fig. 16. LDM along with the measurement data collected at 60 GHz in Riyadh City. The measurements were conducted in a dusty and sunny sky with a visibility of 3 km in 32 °C day-time weather ($h_{tx} = 8.5$ m and $h_{rx} = 3.5$ m).

D2D communications ($h_{tx} = 1.9$ m and $h_{rx} = 1.9$ m). The disparaging effects observed on the PLE values for the case of NLOS are due to the fact that the receiving antenna horn was turned in all directions during the NLOS measurements, in order to capture all significant angle of arrivals at each measurement location. This means that the receiving antenna horn was turned away from the sun rays in most cases during NLOS signal detections. In contrast, the antenna horn was always directed toward the sun rays during the LOS measurement. The level of interactions between solar radio emissions and the propagating waves during the day-time measurements is reflected in the amount of PLE degradations recorded in each case compared with the night-time counterpart.

The higher PLE values observed in hot and sunny sky are mostly attributable to the interference from solar radio emissions arising from the intense solar radiation that characterizes summer afternoon in Riyadh City [25], which causes a decrease in CNR at the input of receiving antennas [18], [21], when compared with the night-time CNR. This translates into a corresponding coverage reduction in hot and sunny sky with intense solar radiations, compared with cool weather at night, when deploying 60 GHz radio outdoors. It was noticed, however, that acceptable signal levels were received at TX–RX separation distances suitable for pico-cell and microcell deployments for both indoor and outdoor scenarios, despite these issues. For example, acceptable signal levels were detected at the TX–RX distance of 134 m in the indoor long hallway measurements shown in Fig. 6, and this range can, in fact, be extended further if the TX and RX antennas were always best pointed to one another using an

Authorized licensed use limited to: Hamad Bin Khalifa University. Downloaded on December 01,2020 at 16:40:53 UTC from IEEE Xplore. Restrictions apply.

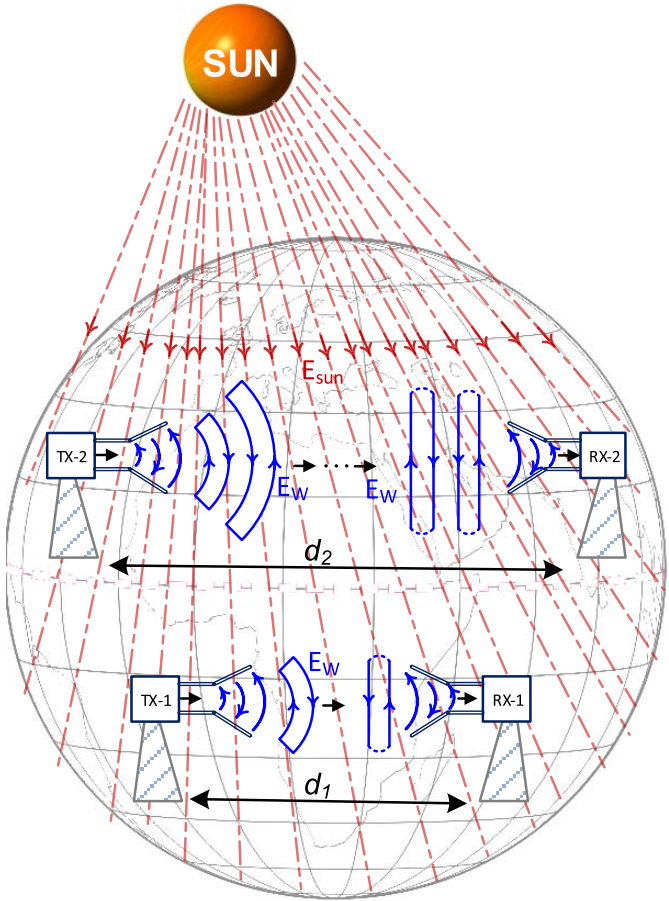

Fig. 17. Electric field incident upon Earth's surface from the Sun, $E_{\text{sun}}$, interacting with the electric field of the propagating waves, $E_w$. $E_{\text{sun}}$: broken lines with arrows. $E_w$: solid curves with arrows.

automatic tracking device. In the outdoor measurements conducted in the rare vegetation area shown in Fig. 3, acceptable signal levels were received at the TX–RX separation distance of up to 110 m in a cool favorable weather in an urban area with rare vegetation, as shown in Fig. 13. Measurements taken in the same neighborhood with some vegetation, however, show reduced coverage as expected, as illustrated in the results in Figs. 14 and 15. During the outdoor measurements, it was also noticed that the signal levels at each TX–RX location were enhanced whenever both directional antennas were best pointed at each other. This emphasizes the importance of beamforming technologies in 5G cellular systems. The results presented in Figs. 14 and 15 reveal that it is possible to deploy 60 GHz radios in the future 5G cellular networks in hilly and dense urban areas with extreme temperatures such as those encountered in the gulf regions, both in indoor and outdoor environments. Good communication links were established in sunny sky during the day and at night, through the use of highly directional steerable horn antennas at both the TX and the RX. Finally, measurement data were collected to assess the combined effects of dusty and sunny weather on the path loss, and the results are shown in Fig. 16. The path loss data shown in Fig. 16 were obtained in a 30 °C sunny and dusty sky, with visibility just at 3 km, for backhaul links ($h_{\text{tx}} = 8.5$ m and $h_{\text{rx}} = 3.5$ m). Comparing the LOS data in Fig. 16 with the counterpart LOS results in Fig. 9 obtained in a 42 °C sunny and clear weather in the same environment and using the same TX and RX heights, it is observed that lower PLE values are obtained in the dusty weather for LOS links. This is due to the fact that dust particles in the sky far above the TX and RX shielded the measurement equipment from the incoming sun rays, reducing the effects of solar radio emissions observed during signal detections in a dusty sky compared with a clear sky. For the NLOS case, the presence of dust particles and flying objects associated with dust/sandstorms (in the vicinity of the TX and RX) technically represents extra scatterers, which degrade the NLOS PLE values as shown in Figs. 9 and 16.

In summary, the effects of solar radio emissions observed in our measurements can be explained from the physics of radio waves as follows. At the operating frequency, let $E_{\text{sun}}$ denote the electric field incident upon Earth's surface from the Sun, and let $E_w$ denote the electric field of the propagating wave as shown in Fig. 17. It can be observed from Fig. 17 that $E_{\text{sun}}$ will interact with $E_w$ as the wave propagates from the TX to the RX, resulting in higher losses (or lower CNR at the RX), in the presence of solar radiation compared with night-time measurements. The level of such interactions depends on the intensity of the solar activities in an area. For example, it is known that in clear weather where the media through which solar power is radiated to the earth is only filled with air, the magnitude of the incident $E_{\text{sun}}$ is: $|E_{\text{sun}}| = (2 \times 377)^{(1/2)} \sqrt{S_{av}}$, where $S_{av}$ is the solar power density experienced in the area, in kW/m² [34]. Therefore, once $|E_{\text{sun}}|$ is known, an exact interaction between $E_{\text{sun}}$ and $E_w$ can be developed based on the angle subtended by the sun. We do not currently have tools to measure $S_{av}$ or $|E_{\text{sun}}|$ in this paper. This can be explored in the future works.

Obviously, the interactions between $E_{\text{sun}}$ and $E_w$ when waves propagate over the distance $d_2$ from TX to RX, as shown in Fig. 17, will be more pronounced compared with the case when the waves propagate over the distance $d_1$, if $d_2 > d_1$. This helps to explain the distance-dependent nature of the effects of solar radio emission on the path loss as observed in our measurements. Also, the orientation of the antennas used in the measurements with respect to the incident $E_{\text{sun}}$ will affect the extent of these interactions for a given TX–RX separation distance. This helps to explain the lower effects observed in our measurement data for the NLOS cases compared with the counterpart LOS measurements taken over the same TX–RX separation distance. Finally, it should be noted that the results presented here are only initial guides for engineers to note what might be the range of typical variations when making path loss budgets for night-time and sunny weather operations. Our future works in this area will conduct similar measurements at the 38 GHz mmWave bands, the results of which will hopefully shed more lights on the effects of solar radio emissions on mmWave signals.

## V. CONCLUSION

In this paper, we have presented the first comprehensive analysis of the effects of solar radio emissions in the 6634 IEEE TRANSACTIONS ON ANTENNAS AND PROPAGATION, VOL. 65, NO. 12, DECEMBER 2017terrestrial wireless communications system at 60 GHz bands. Large-scale propagation path loss models are presented at 60 GHz that incorporate the effect of solar radio emissions on the PLE values. The models were developed based on real-field power measurement campaigns that were conducted during the day and at night in various LOS and NLOS outdoor urban environments in Riyadh City. It is shown that roughly 9.0%–15.6% higher PLE values were obtained in sunny summer afternoons compared with their night-time counterparts, for LOS measurements modeling access and backhaul links as well as D2D communications. This translates into a corresponding decrease in 60 GHz radio coverage in hot and sunny weather during the day. The empirical data are closely corroborated by analytical estimates presented on the cumulative average CNR degradation due to solar radio emissions. These results will help 5G cellular network planners to prepare appropriate link budgets when deploying 60 GHz radios outdoors in extremely hot and sunny weathers, such as those encountered in the gulf region where day-time temperature could reach 42 °C or more. It is also shown, however, that good communication links can be established at 60 GHz at the TX–RX distance of up to 134 m indoors, and up to 110 m outdoors, in a hilly and dense urban area. This range can, in fact, be further extended if the antennas are best pointed to each other. Therefore, it is possible to deploy 60 GHz radio for short-range backhaul/access services as well as for D2D communications in the future 5G networks. In our future works, we plan to conduct measurements at 38 GHz in an effort to further characterize the effects of solar radio emissions at other mmWave bands targeted by 5G cellular system developers.

## References

[1] T. S. Rappaport, R. W. Heath, Jr., R. C. Daniels, and J. N. Murdock, *Millimeter Wave Wireless Communications*. Englewood Cliffs, NJ, USA: Prentice-Hall, 2015.

[2] A. I. Sulyman, A. Alwarafy, H. E. Seleem, K. Humadi, and A. Alsanie, "Path loss channel models for 5G cellular communications in Riyadh city at 60 GHz," in *Proc. IEEE Int. Conf. Commun. (IEEE-ICC)*, May 2016, pp. 1–6.

[3] A. M. Hammoudeh and G. Allen, "Millimetric wavelengths radiowave propagation for line-of-sight indoor microcellular mobile communications," *IEEE Trans. Veh. Technol.*, vol. 44, no. 3, pp. 449–460, Aug. 1995.

[4] R. J. Weiler, M. Peter, W. Keusgen, and M. Wisotzki, "Measuring the busy urban 60 GHz outdoor access radio channel," in *Proc. IEEE Int. Conf. Ultra-Wideband (ICUWB)*, Sep. 2014, pp. 166–170.

[5] R. J. Weiler, M. Peter, W. Keusgen, H. Shimodaira, K. T. Gia, and K. Sakaguchi, "Outdoor millimeter-wave access for heterogeneous networks—Path loss and system performance," in *Proc. IEEE PIMRC*, Sep. 2014, pp. 2189–2193.

[6] A. I. Sulyman, A. T. Nassar, M. K. Samimi, G. R. MacCartney, Jr., T. S. Rappaport, and A. Alsanie, "Radio propagation path loss models for 5G cellular networks in the 28 GHz and 38 GHz millimeter-wave bands," *IEEE Commun. Mag.*, vol. 52, no. 9, pp. 78–86, Sep. 2014.

[7] J. Wells, "Faster than fiber: The future of multi-G/s wireless," *IEEE Microw. Mag.*, vol. 10, no. 3, pp. 104–112, May 2009.

[8] *Narrowband IoT (NB-IoT)*, document RP-151621, RAN Meeting, 3GPP Release 13, Sep. 2015. [Online]. Available: http://www.3gpp.org/news-events/3gpp-news/1733-niot

[9] *IoT-A Architectural Reference Model*, document 257521, accessed: Jul. 2016. [Online]. Available: http://www.iot-a.eu/public

[10] S. Geng, J. Kivinen, X. Zhao, and P. Vainikainen, "Millimeter-wave propagation channel characterization for short-range wireless communications," *IEEE Trans. Veh. Technol.*, vol. 58, no. 1, pp. 3–13, Jan. 2009.

[11] M.-S. Choi, G. Grosskopf, and D. Rohde, "Statistical characteristics of 60 GHz wideband indoor propagation channel," in *Proc. IEEE 16th Int. Symp. Pers., Indoor Mobile Radio Commun. (PIMRC)*, vol. 1. Sep. 2005, pp. 599–603.

[12] R. Mudumbai, S. K. Singh, and U. Madhow, "Medium access control for 60 GHz outdoor mesh networks with highly directional links," in *Proc. IEEE INFOCOM*, Apr. 2009, pp. 2871–2875.

[13] A. R. Tharek and J. P. McGeehan, "Outdoor propagation measurements in the millimetre wave band at 60 GHz," in *Proc. Military Microw.*, vol. 1. 1988, pp. 43–48.

[14] E. A. Grindrod and A. Hammoudeh, "Performance characterisation of millimetre wave mobile radio systems in forests," in *Proc. 10th Int. Conf. Antennas Propag.*, vol. 2. Apr. 1997, pp. 391–396.

[15] E. Ben-Dor, T. S. Rappaport, Y. Qiao, and S. J. Lauffenburger, "Millimeter-wave 60 GHz outdoor and vehicle AOA propagation measurements using a broadband channel sounder," in *Proc. IEEE Global Telecommun. Conf. (GLOBECOM)*, Dec. 2011, pp. 1–6.

[16] P. F. M. Smulders and L. M. Correia, "Characterisation of propagation in 60 GHz radio channels," *Electron. Commun. Eng. J.*, vol. 9, no. 2, pp. 73–80, Apr. 1997.

[17] E. J. Violette, R. H. Espeland, R. O. DeBolt, and F. K. Schwering, "Millimeter-wave propagation at street level in an urban environment," *IEEE Trans. Geosci. Remote Sens.*, vol. GRS-26, no. 3, pp. 368–380, May 1988.

[18] J. A. Kennewell, "Solar radio interference to satellite downlinks," in *Proc. 6th Int. Conf. Antennas Propag. (ICAP)*, Apr. 1989, pp. 334–339.

[19] M. Ibnkahla, Q. M. Rahman, A. I. Sulyman, H. A. Al-Asady, J. Yuan, and A. Safwat, "High-speed satellite mobile communications: Technologies and challenges," *Proc. IEEE*, vol. 92, no. 2, pp. 312–339, Feb. 2004.

[20] L. J. Lanzerotti, D. E. Gary, D. J. Thomson, and C. G. Maclennan, "Solar radio burst event (6 April 2001) and noise in wireless communications systems," *Bell Labs Technol. J.*, vol. 7, no. 1, pp. 159–163, 2002.

[21] J. J. Carr, *Microwave and Wireless Communications Technology*. Newton, MA, USA: Butterworth-Heinemann, 1996.

[22] Lake Travers, ON, Canada. *The Algonquin Radio Observatory*. Accessed: Sep. 2016. [Online]. Available: http://www.arocanada.com/

[23] A. I. Sulyman, A. Alwarafy, G. R. MacCartney, M. K. Samimi, T. S. Rappaport, and A. Alsanie, "Directional radio propagation path loss models for millimeter-wave wireless networks in the 28-, 60-, and 73-GHz bands," *IEEE Trans. Wireless Commun.*, vol. 15, no. 10, pp. 6939–6947, Aug. 2016.

[24] J. F. Kenney and E. S. Keeping, *Mathematics of Statistics*. Princeton, NJ, USA: Van Nostrand, 1965.

[25] E. Zell *et al.*, "Assessment of solar radiation resources in Saudi Arabia," *Solar Energy*, vol. 119, pp. 422–438, Sep. 2015.

[26] D. K. Kilcoyne *et al.*, "Link adaptation for mitigating earth-to-space propagation effects on the NASA scan testbed," in *Proc. IEEE Aerosp. Conf.*, 2016, pp. 1–9.

[27] D. E. Gary, "Solar radio burst effects on wireless systems," in *Proc. IEEE Int. Symp. Electromagn. Compat. (EMC)*, Aug. 2011, pp. 661–664.

[28] S. Pohjolainen, J. Hildebrandt, M. Karlickỳ, A. Magun, and I. Chertok, "Prolonged millimeter-wave radio emission from a solar flare near the limb," *Astron. Astrophys.*, vol. 396, no. 2, pp. 683–692, 2002.

[29] Y. Zhao, Y. Li, H. Zhang, N. Ge, and J. Lu, "Fundamental tradeoffs on energy-aware D2D communication underlaying cellular networks: A dynamic graph approach," *IEEE J. Sel. Areas Commun.*, vol. 34, no. 4, pp. 864–882, Apr. 2016.

[30] E. Franke, "Effects of solar, Galactic and man-made noise on UHF SATCOM operation," in *Proc. Military Commun. Conf. (MILCOM)*, vol. 1. 1996, pp. 29–36.

[31] F. Cleveland, W. Malcolm, D. E. Nordell, and J. Zirker, "Solar effects on communications," *IEEE Trans. Power Del.*, vol. 7, no. 2, pp. 460–468, Apr. 1992.

[32] S. Ito, H. Fukuchi, C. Ohuchi, H. Hirano, and I. Ono, "Statistics of rain attenuation and other environmental effects associated with the BSE satellite down-link at 12 GHz in Japan," *IEEE Trans. Broadcast.*, vol. BC-28, no. 4, pp. 131–138, Dec. 1982.

[33] M. Sumathi, R. Ranjan, R. K. Singh, and P. Kumar, "Performance analysis of sun sensors for satellite systems," in *Proc. Int. Conf. Adv. Electron. Syst. (ICAES)*, 2013, pp. 10–14.

[34] F. T. Ulaby and U. Ravaioli, *Fundamentals of Applied Electromagnetics*, 7th ed. Hoboken, NJ, USA: Pearson Education, 2015.Authorized licensed use limited to: Hamad Bin Khalifa University. Downloaded on December 01,2020 at 16:40:53 UTC from IEEE Xplore. Restrictions apply.

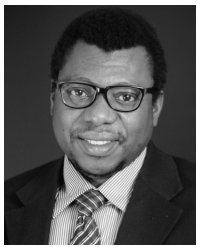

**Ahmed Iyanda Sulyman** (SM'09) received the Ph.D. degree from the Department of Electrical and Computer Engineering, Queen's University, Kingston, ON, Canada, in 2006.

He was a Teaching Fellow with Queen's University from 2004 to 2006, a Post-Doctoral Fellow with the Royal Military College of Canada, Kingston, from 2007 to 2009, and an Assistant/Associate Professor with King Saud University, Riyadh, Saudi Arabia, from 2009 to 2016. He is currently an Associate Professor with the Department of Computer, Electrical, and Software Engineering, College of Engineering, Embry-Riddle Aeronautical University at Prescott, Prescott, AZ, USA. He has authored over 70 technical papers, six book chapters, and a book *Nonlinear MIMO Communication Channels* (Saarbrucken, Deutschland/Germany: LAP LAMBERT Academic Publishing, CRC Press, 2012). His current research interests include wireless communications and networks, with most recent contributions in the areas of millimeter-wave 5G cellular technologies and the Internet of Things.

Dr. Sulyman has served as the session chair and a technical program committee member for many top-tier IEEE conferences, including the 2016 IEEE International Conference on Communications.

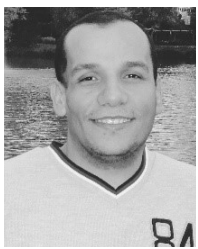

**Hussein Seleem** (S'08) received the B.Sc. (Hons.) and M.Sc. degrees in electronics engineering and electrical communications from Tanta University, Tanta, Egypt, in 2001 and 2006, respectively, and the Ph.D. degree in electrical engineering from King Saud University (KSU), Riyadh, Saudi Arabia, in 2017.

Since 2001, he has been with the Electronics Engineering and Electrical Communications Department, Tanta University, where he is currently a Teaching Assistant. He was a Researcher with the Prince Sultan Advanced Technology Research Institute, KSU, in 2010, where he was mainly involved in optical and wireless communication projects. In 2012, he joined the Electrical Engineering Department, KSU, as a Researcher/Ph.D. Student, where he was involved in the area of wireless communications. His current research interests include signal processing for wireless and optical communications.

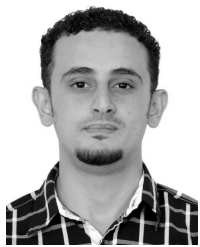

**Abdulmalik Alwarafy** received the B.S. degree in electrical engineering with a minor in communication from Ibb University, Ibb, Yemen, in 2009, and the M.Sc. degree in electrical engineering with a major in communications from King Saud University, Riyadh, Saudi Arabia, in 2015, where he is currently pursuing the Ph.D. degree in electrical engineering.

His current research interests include millimeter-wave measurements and propagation channel modeling and analysis for the fifth-generation cellular mobile communications.

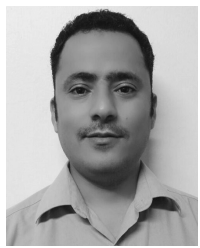

**Khaled M. Humadi** received the B.S. degree in electrical engineering from Ibb University, Ibb, Yemen, in 2010, and the M.Sc. degree (Hons.) in electrical engineering from King Saud University, Riyadh, Saudi Arabia, in 2015, where he is currently pursuing the Ph.D. degree in electrical engineering.

His current research interests include the area of wireless communications theory, including massive multiple input–multiple output systems, spatial modulation, and millimeter-wave technologies for the fifth-generation cellular systems.

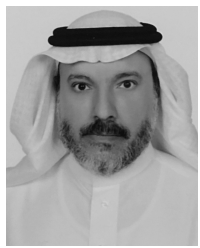

**Abdulhameed Alsanie** received the B.Sc. (Hons.) and M.Sc. degrees in electrical engineering from King Saud University, Riyadh, Saudi Arabia, in 1983 and 1987, respectively, and the Ph.D. degree in electrical engineering from Syracuse University, Syracuse, NY, USA, in 1992.

He is currently an Associate Professor and the Head of the Electrical Engineering Department, King Saud University. His current research interests include wireless communications with a focus on multiple input–multiple output wireless systems, space time codes, and cooperative wireless systems.